\documentclass[a4paper, twocolumn]{article}

\usepackage[utf8]{inputenc}
\usepackage[T1]{fontenc}

\usepackage{graphicx}

\usepackage{amsmath,amsfonts,amssymb}
\usepackage{bm}
\usepackage{extarrows}  
\usepackage{chemarrow}  

\usepackage{dcolumn}
\usepackage{booktabs} 
\usepackage{tocbibind} 
\usepackage{csquotes}
\usepackage{appendix}
\usepackage{mdframed}
\usepackage{caption}
\usepackage{breakcites} 
\usepackage[nodayofweek]{datetime}
\newdateformat{mydate}{\twodigit{\THEDAY}{ }\monthname[\THEMONTH]{ } \THEYEAR}

\usepackage{enumitem} 

\usepackage{etoolbox}
\patchcmd{\chapter}{\thispagestyle{plain}}{\thispagestyle{fancy}}{}{}

\usepackage{etoolbox} 

\newcommand{\eps}{\text{\usefont{OML}{cmr}{m}{n}\symbol{15}}}

\newcommand{\taue}{\ensuremath{\tau_\epsilon}}
\newcommand{\taus}{\ensuremath{\tau_\sigma}}

\newtoggle{bibnew}
\toggletrue{bibnew}
\togglefalse{bibnew}

{

  \usepackage{helvet}
    \usepackage{fourier}
    \usepackage{microtype}
    \usepackage{multirow}
    
\newcommand{\graphicsfolder}{.}

\begin{document}

\title{Spring-damper equivalents of the fractional, poroelastic, and poroviscoelastic models for elastography}

\author{Sverre Holm\\University of Oslo}

\maketitle

\begin{abstract}
In MR elastography it is common to use an elastic model for the tissue's response in order to properly interpret the results. More complex models such as viscoelastic, fractional viscoelastic, poroelastic, or poroviscoelastic ones are  also used. These models appear at first sight to be very different, but here it is shown that they all may be expressed in terms of elementary viscoelastic models. 

For a medium expressed with fractional models, many elementary spring--damper combinations  are added, each of them weighted according to a long-tailed distribution, hinting at a fractional distribution of time constants or relaxation frequencies. This may open up for a  more physical interpretation of the fractional models. 

The shear wave component of the poroelastic model is shown to be modeled exactly by  a three-component Zener model. The extended poroviscoelastic model is found to be equivalent to what is called a non-standard four-parameter model. Accordingly, the large number of parameters in the porous models can be reduced to the same number as in their viscoelastic equivalents. As long as the individual displacements from the solid and fluid parts cannot be measured individually the main use of the poro(visco)elastic models is therefore as a physics based method for determining parameters in a viscoelastic model.
\end{abstract}


\section{Introduction}
In modeling of data from MR elastography, it is common to use a simple elastic model for the medium. This is the case even for ultrasound elastography. For more accurate modeling, there are three families of models that are used. These are the linear viscoelastic models such as the Kelvin-Voigt and Zener models, the fractional extensions of these models, and poroelastic models based on the theory of Biot. 

The linear viscoelastic models are among those that have been used for fitting frequency dependency of shear wave data from MR elastography in the  brain \cite{klatt2007noninvasive}. The fractional Kelvin-Voigt model was fitted to breast MR elastography data in \cite{Sinkus2007} and also analyzed in \cite{Holm2010} and compared to other models for elastography  in \cite{zhang2016estimation}.

It may be argued that these single-phase models are too simplistic and that in  tissue a bi-phasic model which distinguishes between the solid and liquid phases would be more accurate. This potential has already been demonstrated in models of the quasi-static biomechanics of hydrocephalus \cite{nagashima1987biomechanics} and of infusion-induced swelling in the brain \cite{basser1992interstitial}. The poroelastic model has also been used for elastography. In \cite{konofagou2001poroelastography} ultrasound elastography was modeled with either an elastic model (i.e.~without viscosity) or a simplified poroelastic model which depended on porosity, composite density and fluid density. Similarly \cite{perrinez2009modeling} evaluated the full poroelastic model vs.~an elastic one for MR elastography. In \cite{mcgarry2015suitability} they went one step further and compared the poroleastic model with a viscoelastic one with a complex shear modulus. They also tried to use these models for inversion and observed that the viscoelastic model produced better reconstructions at 50 Hz while the poroelastic model was superior at 1 Hz. 

The challenge in making a reconstruction algorithm based on the poroelastic model is the need for capturing the displacement of the solid and the fluid independently. Chapter 5 in \cite{hirsch2016magnetic} states that due to the voxel size of MR elastography, it cannot detect properties of individual pores, but only sees a homogeneous effective medium. Further, since it is sensitive to signals from hydrogen atoms in the voxel, one cannot separate the signal from the solid and the fluid, even if the solid should contain fluid which is considered not to be in the pores. Likewise ultrasound is scattered from the soft tissue which is mainly part of the solid matrix. Ultrasound will therefore produce a strain estimate in the solid matrix which is only indirectly influenced by the wave in the fluid \cite{konofagou2001poroelastography}. In this paper, the goal is however not to develop a poroelastic model for reconstruction of the MR elastography image, but only to discuss it in the framework of explaining variations in parameters due to physiological and physical parameters.

The claim of this paper is that these models are much more similar than they appear to be. The fractional models can be developed as sums of ordinary viscoelastic elements weighted in a particular way. The fractional model has its strength in that it gives a parsimonious description of the phenomenon, i.e.~one with a minimal number of parameters, especially when power-law variation in frequency is observed. But it is not fundamentally different. Likewise the poroelastic model for the wave mode of interest for elastography, the shear wave, can be described in terms of standard viscoelastic models. In this case it is the viscoelastic model which requires the smallest number of parameters. The strength of the poroelastic formulation is that it gives a way of finding how these parameters depend on physical and even physiological parameters. 

The linear viscoelastic model is therefore first generalized in Sec.~\ref{sec:Viscoelastic} from the simple two- and three-component models to chains of spring-damper elements described by time- and frequency-spectral functions. These functions are used to show that the fractional viscoelastic models in Sec.~\ref{sec:Fractional} can be described as sums of ordinary viscoelastic models. A surprising result is that the weighting in the sum follows a  long-tailed distribution reminiscent of a fractal. 

The shear wave solution of the poroelastic theory is then developed in Sec.~\ref{sec:Poroelastic} both from the original formulation of Biot \cite{biot1956theoryI} and also from that of Stoll \cite{stoll1977acoustic}. Somewhat unexpectedly, it is found that there is a one to one correspondence between the poroelastic shear wave response and that of a simple spring-dashpot network.

\section{Linear viscoelastic models}
\label{sec:Viscoelastic}
The linear viscoelastic model is expressed in three different ways. The different ways are needed in order to generalize from the simple two and three-component models (e.g.~Kelvin-Voigt and Zener) to higher order models. 

In order to illustrate the similarities between models it is sufficient here to express the models in one dimension, although three dimensions are really needed for a complete shear wave description. This section, as well as Sec.~\ref{sec:Fractional} builds to a large degree on \cite{Mainardi2010}.

\subsection{Three descriptions}
\label{sec:threeDescriptions}

In linear viscoelasticity there are three different ways of describing the medium's response. The first is the hereditary model of Boltzmann \cite{boltzmann1876theorie,markovitz1977boltzmann} where the constitutive equation is a convolution integral \cite{Mainardi2010} (ch.~2):
\begin{align}
\sigma(t) =  G(t)* \frac{\partial\epsilon(t)}{\partial t}.
\label{eq:convConstPhysical}
\end{align}
The kernel $G(t)$ is called the relaxation modulus and represents a fading memory, i.e. one where changes in the past have less effect now than more recent changes.  In order to ensure causality, the kernel, $G(t)$,  has to be zero for non-negative time. The relaxation modulus is the strain response of the system to a step excitation in strain.

\begin{figure}[t]
 \centering
 \includegraphics[width=0.5\columnwidth]{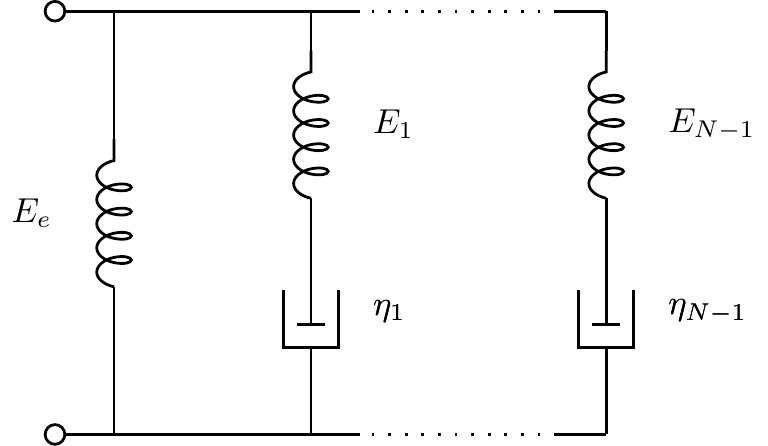}
 \caption{Maxwell-Wiechert model consisting of parallel networks of spring-damper models}
 \label{fig:Maxwell-Wiechert}
\end{figure}

The second description is a linear differential equation between stress and strain with constant coefficients:  
\begin{align}
	\left[1 + \sum_{k=1}^{p} a_k \frac{\partial^{k}}{\partial t^{k}}\right] \sigma(t)  = \left[E_e + \sum_{k=1}^{q} b_k \frac{\partial^{k}}{\partial t^{k}}\right] \epsilon(t).
\label{eq:highOrder}
\end{align}
The third description is in the form of a relaxation spectrum or Prony expansion:
\begin{equation}
G(t)=G_e + G_\tau + G_-\delta(t),
\enspace G_\tau(t) = \sum_{n=0}^{N-1} E_n \exp(-t/\tau_n).
\label{eq:RelaxationSpectrum}
\end{equation}
A physically realizable model is obtained if $G(t)$ is modeled by a parallel network of pairs of springs and dashpots in series where the spring constants are $E_n$ and the viscosity of the dashpots are $\eta_n = \tau_n E_n$. Furthermore, both the spring constants and the viscosities are non-negative. This is the Maxwell-Wiechert model of Fig.~\ref{fig:Maxwell-Wiechert}. The left-hand spring leads to the constant $G_e=E_e$, the equilibrium modulus, and if there is a dashpot directly across the terminals (e.g. if $E_1=0$), the response will have an impulse at time zero as well given by $G_-$. This is the case in the Kelvin-Voigt model which will soon be discussed.

Not all the coefficient sets of \eqref{eq:highOrder} will lead to a fading memory model, so the linear differential equation is the most general model. Also, a model described as a relaxation spectrum, \eqref{eq:RelaxationSpectrum}, can always be described with a linear differential equation, but not vice versa. In addition, a relaxation spectrum model, \eqref{eq:RelaxationSpectrum}, always results in a fading memory model as it consists of sums of positively weighted falling exponentials. 

The frequency domain equivalent of \eqref{eq:highOrder} is also useful:
\begin{equation}
 E(\omega)  =  \frac{\sigma(\omega)}{\epsilon(\omega)} =
\frac{E_e + \sum_{k=1}^q b_k (i\omega)^k}
{1 + \sum_{k=1}^{p} a_k (i\omega)^k}
\label{eq:FreqDomain}
\end{equation}
The dynamic modulus, $E(\omega)$, is often called $G(\omega)$ in elastography, but since $G$ here means the relaxation modulus in this paper, $E$ is used instead.

\subsection{Elementary linear viscoelastic models}
\label{sec:ElementaryViscoleastic}

\begin{figure}[t]
 \centering
 \includegraphics[width=0.5\columnwidth]{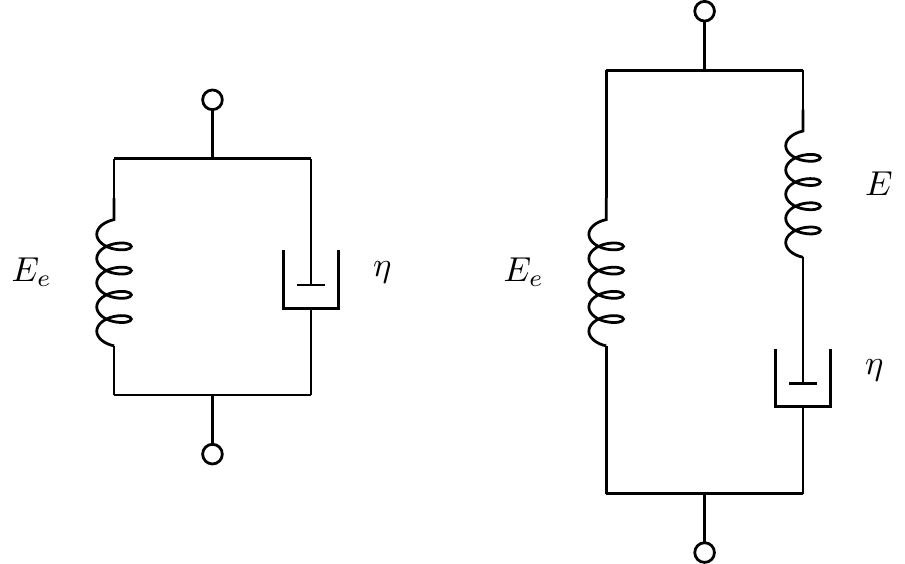}
 \caption{Kelvin-Voigt (left) and Zener models. As shown in this paper, the three-component Zener model also models the shear wave in the Biot poroelastic model}
 \label{fig:Kelvin-Voigt_Zener}
\end{figure}

The two simplest viscoelastic models are the Kelvin-Voigt and the Zener models as shown in Fig.~\ref{fig:Kelvin-Voigt_Zener}. The Kelvin-Voigt model is found by just keeping $E_e$ and $\eta_1$ in Fig.~\ref{fig:Maxwell-Wiechert}. The linear differential equation is
\begin{equation}
\sigma(t) = E_e \epsilon(t) + \eta \frac{\partial \epsilon(t)}{\partial t}, 
\end{equation}
and the relaxation modulus is:
\begin{equation}
G(t) = E_e  + \eta  \delta(t).
\label{eq:KVRelaxationModulus}
\end{equation}
Thus $p=0$, $q=1$, $a_1=0$, and $b_1=\eta$ in \eqref{eq:highOrder} and \eqref{eq:FreqDomain}. Further $G_e=E_e$, $G_-=\eta$, and $G_\tau(t)=0$ in \eqref{eq:RelaxationSpectrum}.
The dynamic modulus is:
\begin{equation}
E(\omega) = E_e + i \omega \eta = E_e (1+ i \omega \tau_\epsilon)
\label{eq:KV-dynamicModulus}
\end{equation}
where $\tau_\epsilon = \eta/E_e$. The even simpler elastic model, which is often the reference in elastography as noted in the Introduction, is obtained by setting  the viscosity, $\eta$, to 0.

The Zener model adds one more term to the linear differential equation of the Kelvin-Voigt model:
\begin{equation}
\sigma(t) + a_1 \frac{\partial \sigma(t)}{\partial t} = E_e \epsilon(t) + \eta \frac{\partial \epsilon(t)}{\partial t},
\end{equation}
and the relaxation modulus is:
\begin{equation}
G(t) = E_e + E_e(\frac{\tau_\epsilon}{\tau_\sigma}-1) e^{-t/\tau_\sigma}.
\end{equation}
Compared to the Kelvin-Voigt model, the effect of the second spring is to `soften' the impulse in \eqref{eq:KVRelaxationModulus} into a falling exponential. 
In this model $p=q=1$ and $b_1=\eta$  in \eqref{eq:highOrder} and \eqref{eq:FreqDomain}. The time constants are $\tau_\epsilon$ as in the Kelvin-Voigt model and $\tau_\sigma = a_1$. 
The parameters of the model relate to the physical parameters of Fig.~\ref{fig:Maxwell-Wiechert} as follows:
%
%
\begin{equation}
\tau_\sigma = \eta/E \le  \tau_\epsilon = \eta/E', \quad \frac{1}{E'} = \frac{1}{E_e} + \frac{1}{E}.
\label{eq:ZenerCondition}
\end{equation}
For this model the dynamic modulus is:
\begin{equation}
E(\omega) =E_e \frac{1+ i \omega \tau_\epsilon}{1+ i  \omega \tau_\sigma}
\label{eq:Zener-dynamicModulus}
\end{equation}
The addition of the extra spring leads to frequency dependent denominator in the dynamic modulus which gives more degrees of freedom in fitting this model to data.

The dynamic modulus can also be included in a dispersion relation when propagating waves are involved. In \cite{Holm2011}, Eqs. (13-15), it is shown that the complex wave number in that case is:
%
%
\begin{equation}
\left( \frac{k}{\omega}\right)^2 = \rho \kappa(\omega) = \frac{\rho}{E(\omega)}
\label{eq:Wavenumber}
\end{equation}
where $\rho$ is the density and $\kappa(\omega)$ is the dynamic compressibility, the inverse of the dynamic modulus. This result is important in the subsequent analysis of the poroelastic model. Note that the roles of $\tau_\epsilon$ and $\tau_\sigma$ are reversed here compared to \cite{Holm2011}. Here the convention of \cite{Mainardi2010} is followed instead.

A slightly rewritten version of the compressibility based on factoring of \eqref{eq:Zener-dynamicModulus} will be needed when analyzing the poroelastic model:
\begin{equation}
\kappa(\omega) = \frac{1}{E(\omega)} = \frac{1}{E_e} \frac{1+ \omega^2 \tau_\epsilon \tau_\sigma - i \omega(\tau_\epsilon-\tau_\sigma)}{1+  \omega^2  \tau_\epsilon^2}
\label{eq:Zener-dynamicModulus2}
\end{equation}

Judging from Fig.~\ref{fig:Maxwell-Wiechert}, a Zener model is characterized by two springs and a damper. These three parameters may be expressed in different ways, and one common way is via the low frequency asymptote of the propagation speed, $c_0$, and the two cross-over frequencies, $\omega_\epsilon$ and $\omega_\sigma$. The two frequencies express approximately where the phase velocity starts rising and where it approaches its asymptotic value $c_\infty = c_0 \sqrt{\omega_\sigma/\omega_\epsilon}$ \cite{Holm2011}.  Taking into consideration that $c_0^2 = E_e/\rho$ and thus depends on both the shear modulus and the density, then a Zener medium will depend on four independent parameters: $E_e$, $\rho$, $\omega_\epsilon$ and $\omega_\sigma$.

\subsection{Time- and frequency-spectral functions}

The continuous generalization of \eqref{eq:RelaxationSpectrum} is given in  \cite{gross1947creep,caputo1971linear, Mainardi2010}:
\begin{equation}
G_\tau(t) = b \int_0^\infty R_\sigma(\tau)e^{-t/\tau} d\tau, 
\end{equation}
where $b=G_\tau(0)$ is a non-negative constant which for the Zener model is $b= E(\tau_\epsilon/\tau_\sigma-1)$ \cite{caputo1971linear}.
$R_\sigma(\tau)$ is a non-negative relaxation spectrum.

There is a corresponding frequency-spectral function which is
\begin{equation}
S_\sigma(\Omega) = b\frac{R_\sigma(1/\Omega)}{\Omega^2} 
\end{equation}
By substituting $\Omega=1/\tau$ so $d\Omega = -d\tau/\tau^2$ this gives
\begin{equation}
G_\tau(t) =  \int_0^\infty S_\sigma(\Omega)e^{-\Omega t} d\Omega.
\end{equation}
The relaxation modulus is, as noted, the stress response of the system to a step excitation in strain. The relaxation spectrum could also have been expressed with the creep response, which is the strain response to  a step excitation in strain. It is denoted by $J(t)$ and plays the same role as $G(t)$ in \eqref{eq:convConstPhysical} if $\sigma(t)$ and $\epsilon(t)$ are interchanged. The time-spectral function is:
\begin{equation}
J_\tau(t) = a \int_0^\infty R_\epsilon(\tau) (1-e^{-t/\tau}) d\tau
\label{eq:TimeSpectralCreep}
\end{equation}
where  $a=J_\tau(\infty)$ is a non-negative constant which for the Zener model is $a= (1/E)(1-\tau_\sigma/\tau_\epsilon)$ \cite{caputo1971linear}. This corresponds to a decomposition in a sum of elementary models in series as shown in Fig.~\ref{fig:MultipleVoigt-Kelvin}. This is the conjugate of the Maxwell-Wiechert model of Fig.~\ref{fig:Maxwell-Wiechert}, i.e.~a different configuration of springs and dash-pots which has the same characteristics  \cite{tschoegl1989phenomenological}. It follows that the conjugate of the Zener model in the right-hand part of  Fig.~\ref{fig:Kelvin-Voigt_Zener} is the top three elements of Fig.~\ref{fig:MultipleVoigt-Kelvin} also.

The frequency-spectral function for the creep is found by a transformation which is analogous to that for the relaxation:
\begin{figure}[t]
 \centering
 \includegraphics[width=0.3\columnwidth]{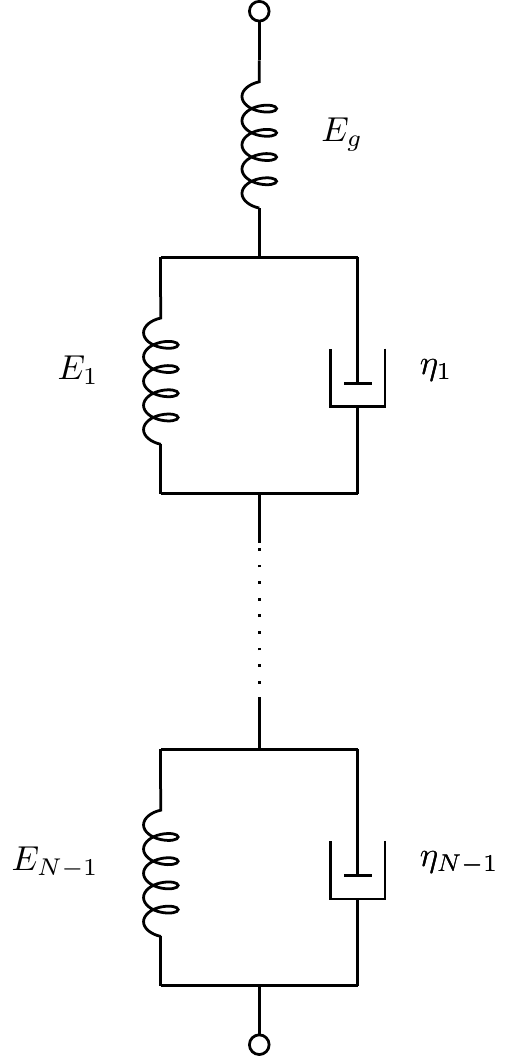}
 \caption{Kelvin model consisting of series networks of spring-damper models. It is the conjugate of the Maxwell-Wiechert model of Fig.~\ref{fig:Maxwell-Wiechert}}
 \label{fig:MultipleVoigt-Kelvin}
\end{figure}
\begin{equation}
S_\epsilon(\Omega) =  a\frac{R_\epsilon(1/\Omega)}{\Omega^2} 
\label{eq:FrequencyTimeCreep}
\end{equation}
and the frequency-spectral function is:
\begin{equation}
J(t) = - \int_0^\infty R_\sigma(\epsilon)(1-e^{-t/\tau}) d\tau
\label{eq:FrequencySpectralCreep}
\end{equation}
This frequency spectral function will be used to find the spring-damper equivalent of the fractional models.

\section{Fractional models}
\label{sec:Fractional}

\begin{figure}[tb]
 \centering
 \includegraphics[width=0.3\columnwidth]{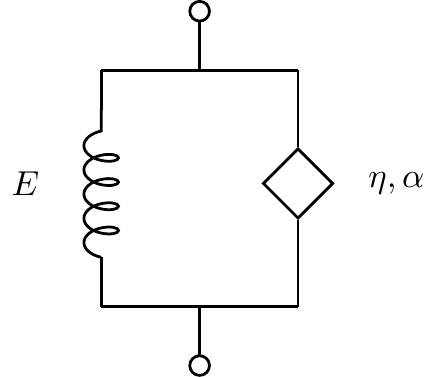}
 \caption{Fractional Kelvin-Voigt model with spring characterized by shear modulus, $E$, and spring-pot given by shear viscosity, $\eta$, and fractional order, $\alpha$}
 \label{fig:fractKV}
\end{figure}

Both the Kelvin-Voigt model and the Zener model can be generalized by introducing non-integer, fractional derivatives of order $\alpha$ in the constitutive equation \eqref{eq:highOrder}. For the Zener model this is:
\begin{align}
	\sigma(t) +\tau_{\sigma}^{\alpha} \frac{\partial^{\alpha}\sigma(t)}{\partial t^{\beta}}  = E \left[\epsilon(t) +\tau_{\epsilon}^{\alpha} \frac{\partial^{\alpha}\epsilon(t)}{\partial t^{\alpha}}\right]
\label{Eq:gZener}
\end{align}
The dynamic modulus is now:
\begin{align}
	E(\omega) = E_e \frac{1+ (i \omega \tau_{\epsilon})^{\alpha}}{1 + (i \omega\tau_{\sigma})^{\alpha}}
\label{Eq:FracZenerDynamicModulus}
\end{align}
while in the simpler Kelvin-Voigt model shown in Fig.~\ref{fig:fractKV}, the constant $\tau_{\sigma}=0$:
\begin{align}
\sigma(t) =
 E \left[\epsilon(t) +\tau^{\alpha} \frac{\partial^{\alpha}}{\partial t^{\alpha}}\epsilon(t) \right].
\label{eq:fractKVConst}
\end{align}
The dynamic modulus is:
\begin{align}
	E(\omega) = E_e \left(1+ (i \omega \tau_{\epsilon})^{\alpha} \right)
\label{Eq:FracKVDynamicModulus}
\end{align}
In the limit as the frequency and/or viscosity is very large the dynamic modulus approaches $E(\omega)  \rightarrow (i \omega \eta)^{\alpha}$. This is equivalent to the case where the spring of Fig.~\ref{fig:fractKV} can be neglected. That seems to be the case often for elastography data \cite{zhang2007congruence, sinkus2017rheological}. 

The relaxation modulus of the fractional Kelvin-Voigt model is expressed by a power law function of time, while for the fractional Zener model it follows a Mittag-Leffler function. That function is a generalization of the exponential function with power-like behavior. The creep compliance of both models also follows a Mittag-Leffler function. That means that the creep time- and -frequency-spectral functions of the two models will be similar also.

\subsection{Fractal time- and frequency-spectral function}

In \cite{gross1947creep,caputo1971linear, Mainardi2010} it has been shown that for the fractional Zener model, the creep time-spectral function of \eqref{eq:TimeSpectralCreep} is:
\begin{equation}
R_\epsilon(\tau) = \frac{1}{\pi \tau} \frac{\sin{\alpha \pi}}{(\tau/\tau_\epsilon)^\alpha + (\tau/\tau_\epsilon)^{-\alpha} + 2 \cos{\alpha \pi}} 
\end{equation}
It was plotted in \cite{caputo1971linear, Mainardi2010} with linear axes and is a decreasing function of $\tau$ for small $\alpha$ and gets a more and more pronounced peak as $\alpha$ approaches 1. For the non-fractional case, $\alpha=1$ it is a delta function at $\tau/\taue=1$ showing that it is equivalent to  a single relaxation process in that case.


Here the function is plotted on a log log scale in Fig.~\ref{fig:TimeSpectral}. That  brings out the properties of this particular  function in a different way than if it is plotted in a linear plot. The log log plot fixes the attention on the asymptotes, rather than the peak, and they are:
\begin{figure}[bt]
	\begin{center}
		\includegraphics[width=0.8\columnwidth]{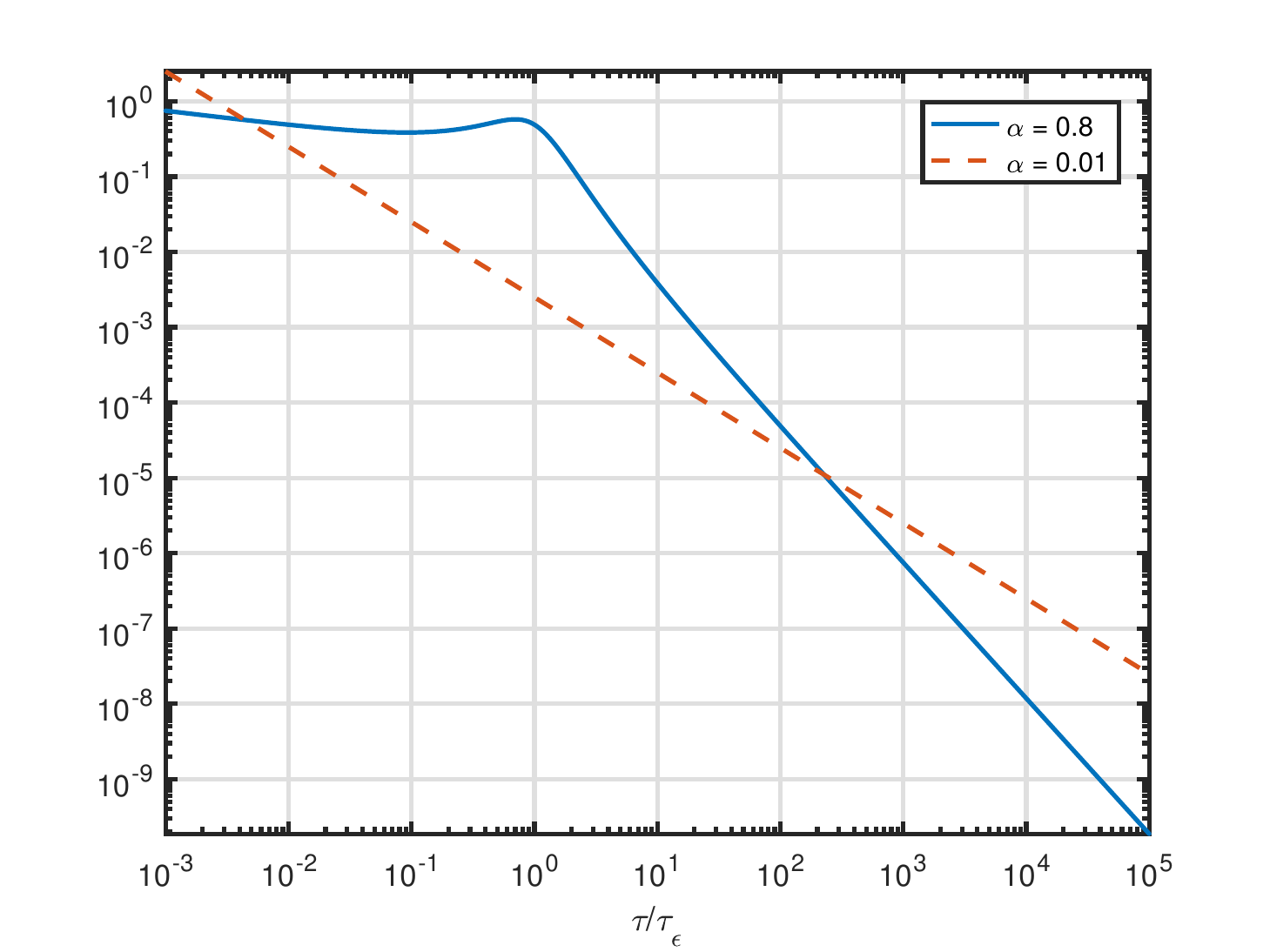}
	\end{center}
	\caption{Time-spectral function for the fractional Zener and Kelvin-Voigt models. Solid line $\alpha=0.8$, dashed line $\alpha=0.01$}
    \label{fig:TimeSpectral}
\end{figure}
\begin{align}
R_\epsilon(\tau) \propto 
   \begin{cases}
        \tau^{\alpha-1}         &  \text{for} \quad \tau/ \taue \ll 1,\\
        \tau^{-\alpha-1}        & \text{for} \quad \tau/\taue \gg 1\\ 
    \end{cases}
	\label{eq:asymptote_TimeSpectral}
\end{align}
Both of them are power laws.  As long as the exponent of the high-frequency tail falls off slower than $\tau^{-2}$, i.e.\ for $\alpha <1$, the variance of the distribution will not exist.
Such long-tailed distributions are scale-invariant  and therefore may indicate fractal properties. This indicates that the distribution of elementary spring-damper models in the medium, as given in Figs.~\ref{fig:Maxwell-Wiechert} and \ref{fig:MultipleVoigt-Kelvin}, may be a multi-fractal with two different fractal orders.

The references above didn't derive the frequency-spectral function, but it can be found by substituting \eqref{eq:FrequencyTimeCreep} and using the normalizing constant $a= J_\tau(\infty) =(1/E)(1-(\tau_\sigma/\tau_\epsilon)^\alpha)$ for the fractional Zener model.  The creep frequency-spectral function of \eqref{eq:FrequencySpectralCreep} is then:
\begin{align}
	S_\epsilon(\Omega) =
	 \frac{1}{\pi E} \frac{(\taue^\alpha- \taus^\alpha)\Omega^{\alpha-1} \sin{\alpha\pi}  }{ (\Omega\taue)^{2\alpha} + 1 +  2(\Omega\taue)^\alpha \cos{\alpha\pi} }.
	\label{eq:distribution_for_alphaisbeta}
\end{align}
This result was first given in \cite{Nasholm2011} in the form  above. There it was derived with a starting point in the multiple relaxation theory of \cite{Nachman1990}. It should be noted that the frequency-spectral function is exactly the same as the time-spectral function except for a scaling factor.

\begin{figure}[bt]
	\begin{center}
		\includegraphics[width=0.8\columnwidth]{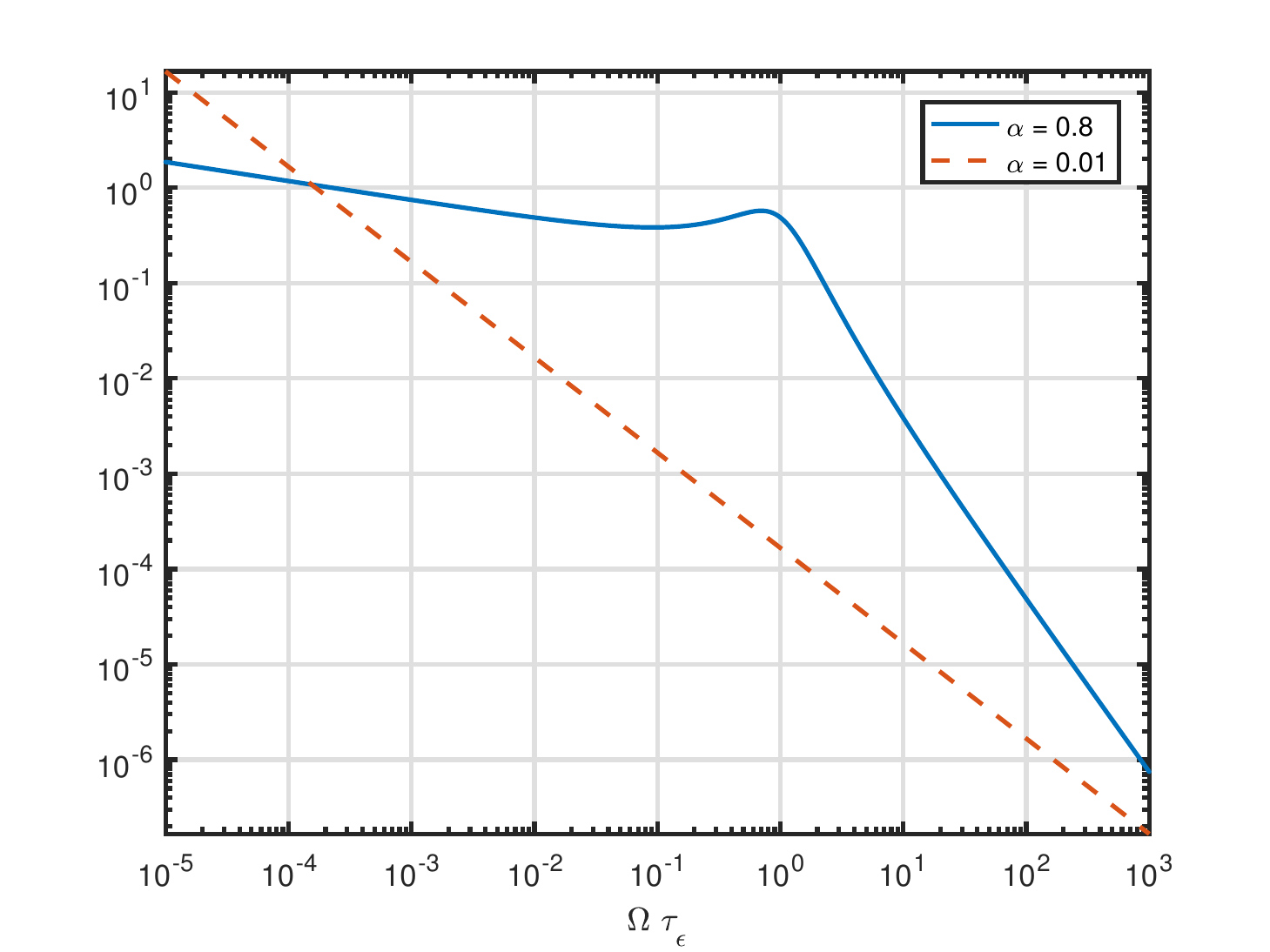}
	\end{center}
	\caption{Frequency-spectral function for the fractional Zener model for $\tau_\sigma = 1000 \tau_\epsilon$. Solid line $\alpha=0.8$ as in \cite{Nasholm2011}, dashed line $\alpha=0.01$. It is also valid for the Kelvin-Voigt model except for an amplitude scaling}
    \label{fig:relax}
\end{figure}

The plot for $\alpha=0.8$ and $\alpha=0.01$ in Fig.\ \ref{fig:relax} shares many properties with the plot of the time-spectral function. The asymptotes are also very similar:
\begin{align}
S_\epsilon(\Omega) \propto 
   \begin{cases}
        \Omega^{\alpha-1}         &  \text{for} \quad \Omega \taue \ll 1,\\
        \Omega^{-\alpha-1}        & \text{for} \quad \Omega \taue \gg 1\\ 
    \end{cases}
	\label{eq:asymptote_Relax}
\end{align}
Interestingly, the relaxation spectrum approaches a single fractal for the limiting case $\alpha \rightarrow 0$, where both the low- and high-frequency parts will approach $\Omega^{-1}$. See the dash-dotted line for $\alpha =\eps = 0.01$ in Fig.\ (\ref{fig:relax}) with asymptotes $\Omega^{-1+\eps}$ for low frequencies and $\Omega^{-1-\eps}$ for high frequencies.

%
%

\subsection{Relationship with hysteresis}
\index{hysteresis}
The results found here have some resemblance with those derived from  hysteresis. Hysteresis is  a hypothetical loss element with a constant phase lag between stress and strain at all frequencies \cite{parker2014real} resulting in attenuation that increases linearly with frequency \cite{stoll1970wave}. The dynamic modulus for hysteresis is:
\begin{align}
E(\omega) = K + i H 
\label{eq:Hysteresis}
\end{align}
where $K$ and $H$ are constants. The model leads to a noncausal model and therefore \cite{parker2014real} is concerned with making a bandlimited approximation where the constants are allowed to vary with frequency, $K(\omega)$ and $H(\omega)$, in order to ensure causality.

By comparing the dynamic moduli, \eqref{eq:Hysteresis} and \eqref{Eq:FracKVDynamicModulus}, it is evident that hysteresis can be viewed as the limiting case as $\alpha\rightarrow 0$ for the fractional Kelvin-Voigt model. Thus an alternative way to ensure causality is to employ a fractional Kelvin-Voigt model with parameter $\alpha=\eps = 0^+$. 

Hysteresis is particularly interesting in the context of fractional  operators because the result for the limiting case $\alpha \rightarrow 0$ of the previous section is similar to that recently found for hysteresis  in \cite{parker2015could}.
There it was shown that the hysteresis model is the same as a sum of relaxation processes weighted with a long-tailed power law $ \Omega^{-1+\eps}$. This parallels our discussion of the asymptotic result of \eqref{eq:asymptote_Relax} for  $\alpha=\eps$ and Parker's result can be interpreted as a special case of \eqref{eq:asymptote_Relax}.

\section{Poroelastic model}
\label{sec:Poroelastic}
The Biot poroelastic model deals with a saturated porous medium with a solid phase and a fluid phase. Wave propagation in such a medium is described by a set of coupled vector wave equations as given in Eq.~(4.2) in \cite{biot1956theoryI}. The variables are the displacement vectors $\mathbf{u}$ and $\mathbf{U}$ for the displacement in the solid and the fluid respectively. That paper assumes that as the fluid moves in the pores, the flow is laminar and that  losses are given by Darcy's law and proportional to the relative velocity, $\partial(\mathbf{u - U})/\partial t$. The theory predicts three solutions, two compressional waves and one shear wave. In a biological porous medium like cancellous bone (bone with a low volume fraction of solid, less than 70\%), all three waves have been detected \cite{hosokawa1997ultrasonic}.


This model is much more complex than the viscoelastic models of the previous sections. It also seems more appropriate for a complex medium like  brain, liver, or bone. Here a surprising exact relationship between the poroelastic and viscoelastic models will be shown for the shear wave solution.

\subsection{Biot's original formulation}

As elastography is only concerned with shear waves, we restrict the analysis here to that mode. Then the following dispersion relation was derived in \cite{biot1956theoryI}, Eqs.~(7.5)--(7.6):
\begin{equation}
\left( \frac{k}{\omega} \right)^2 = \rho (\kappa_r - i \kappa_i),
\label{eq:Biot7.5}
\end{equation}
where
%
\begin{align}
\kappa_r = \frac{1}{\mu} \frac{1+\gamma_{22}\frac{\gamma_{11}\gamma_{22}-\gamma_{12}^2}{(\gamma_{12}+\gamma_{22})^2}\left(\frac{f}{f_c}\right)^2}{1+\left(\frac{\gamma_{22}}{\gamma_{12}+\gamma_{22}} \right)^2\left(\frac{f}{f_c}\right)^2}, 
\kappa_i =  \frac{1}{\mu} \frac{f}{f_c} \frac{\gamma_{12}+\gamma_{22}}{1+\left(\frac{\gamma_{22}}{\gamma_{12}+\gamma_{22}} \right)^2\left(\frac{f}{f_c}\right)^2}
\label{eq:Biot7.6}
\end{align}
This expression depends on three normalized densities:
\begin{equation}
\gamma_{11} = \frac{\rho_{11}}{\rho}, \enspace
\gamma_{22} = \frac{\rho_{22}}{\rho}, \enspace
\gamma_{12} = \frac{\rho_{12}}{\rho},
\end{equation}
and the aggregate or composite density which is
\begin{equation}
\rho = \phi \rho_f + (1-\phi) \rho_s.
\end{equation}
These formulas depend on the porosity, $\phi$, and the fluid and solid densities. The parameter $\rho_{12}$ represents a negative mass coupling density between fluid and solid, $\rho_{11}$ represents the total effective mass of the solid moving in the fluid, and $\rho_{22}$ is the total effective mass of the fluid. There is also a characteristic frequency
\begin{equation}
f_c = \frac{b}{2 \pi (\rho_{12} + \rho_{22})}, \enspace b = \frac{\eta \phi^2}{B}
\end{equation}
where $\eta$ is the fluid viscosity, and $B$ is the permeability. The effective characteristic frequency of \eqref{eq:Biot7.6} is however a lower frequency: 
\begin{equation}
f_c' = f_c \frac{\gamma_{12}+\gamma_{22}}{\gamma_{22}} = \frac{1}{2 \pi}\frac{\eta \phi}{B \alpha \rho_f}
\label{eq:charFreqPrime}
\end{equation}
since $\rho_{22} = \alpha \phi \rho_f$ \cite{berryman1980confirmation} where $\alpha$ is a structure constant related to tortuosity.

Comparison with \eqref{eq:Wavenumber} and \eqref{eq:Zener-dynamicModulus2} shows that \eqref{eq:Biot7.5} and \eqref{eq:Biot7.6} are exactly the same as those of the Zener model. This is a remarkable result which shows that for shear waves, the poroelastic model is also a linear viscoelastic model. What distinguishes it from the ordinary linear viscoelastic models is that it provides a sophisticated way of determining the parameters from physical considerations.

\subsection{Biot-Stoll formulation}

The parameters of the original Biot theory are often considered to be hard to find in practice and the theory has been rewritten in terms of the relative displacement between fluid and solid, $\mathbf{u - U}$, or the volume of fluid that has flowed in or out of an element, $\zeta = \phi \nabla(\mathbf{u - U})$, in combination with the solid displacement $\mathbf{u}$. The material parameters are then  transformed to a new set of parameters.

In that case
Eq.~(16.77) in \cite{hovem2012marine} gives the shear dispersion of the low frequency Biot model. It can also be found from  Eq.~(12) of \cite{stoll1977acoustic}:
\begin{equation}
\left( \frac{k}{\omega} \right)^2 
= \frac{\rho}{\mu} \frac{(\rho_c -\frac{\rho_f^2}{\rho}) - i \frac{\eta }{\omega B}} {\rho_c - i \frac{\eta }{\omega B}}
= \frac{\rho}{\mu} \frac{1 + i \omega \left(\rho_c - \frac{\rho_f^2}{\rho}\right) \frac{B}{\eta}} {1 + i \omega \frac{\rho_c B}{\eta}}
\label{eq:Hovem16.77}
\end{equation}
where the sign of $\omega$ has been reversed compared to the original articles due to a different definition of the Fourier transform than here (see also Appendix A of \cite{holm2014comparison}). The mass coupling density is
\begin{equation}
\rho_c = \alpha \rho_f/\phi
\label{eq:MassCouplingDensity}
\end{equation}
where $\alpha$ is the tortuosity. 
Comparing \eqref{eq:Hovem16.77} to \eqref{eq:Wavenumber} shows that the dynamic modulus is:
\begin{equation}
E(\omega) = \rho \left( \frac{\omega}{k} \right)^2 
= \mu   \frac{1 + i \omega \frac{\rho_c B}{\eta}} {1 + i \omega \left(\rho_c - \frac{\rho_f^2}{\rho}\right) \frac{B}{\eta}} 
\label{eq:BiotShear-dynamicModulus}
\end{equation}
As expected this is also equivalent to that of a Zener model as can be seen by comparison with \eqref{eq:Zener-dynamicModulus}.
The constants when $\omega_\epsilon=1/\tau_\epsilon$ and $\omega_\sigma=1/\tau_\sigma$ are:
\begin{equation}
E_e=\mu_r, \quad \omega_\epsilon =  \frac{\eta}{\rho_c B}, \quad \omega_\sigma = \frac{\omega_\epsilon}{1- \frac{\rho_f^2}{\rho \rho_c}} \ge \omega_\epsilon,  \quad c_0^2=\frac{\mu}{\rho}.  
 \label{eq:BiotParameters}
\end{equation}
%
Insertion of the expression for the mass coupling density, \eqref{eq:MassCouplingDensity}, in the equation for the effective characteristic frequency, \eqref{eq:charFreqPrime}, shows that this frequency also is the same as $\omega_\epsilon/(2 \pi)$.
\begin{table}
\begin{center}
\begin{tabular}{|c|c|l|c|}
     \hline
      \multicolumn{3}{|l|}{\textbf{Biot model parameters}} &  Shear?\\
	\hline 
	\multicolumn{4}{|l|}{\textbf{Bulk properties:}} \\
	\hline
	$\phi$  &[0 ... 1]  &Porosity&  Y \\
	\hline
	$\rho_s$ & [kg/m$^3$] & Solid density & Y \\ 
    \hline

	$\rho_f$ & [kg/m$^3$] & Fluid density & Y \\ 
	\hline 
	$K_s$ &[Pa]  &Bulk modulus, solid  & N \\ 
	\hline 
	$K_f$ &[Pa]  &Bulk modulus, fluid  & N \\ 
	\hline
	\multicolumn{4}{|l|}{\textbf{Fluid parameters:}} \\
	\hline
	$\eta$ &[Pa s]  &Viscosity  & Y \\ 
	\hline 
	$B$ & [m$^2$] & Permeability &Y  \\ 
	\hline 
	$\alpha$& [1 ... 3]   &Tortuosity  &Y  \\ 
	\hline 
	\multicolumn{4}{|l|}{\textbf{Rigid frame response parameters:}} \\
	\hline
	$K_r$&[Pa]  & Bulk modulus& N \\ 
	\hline 
	$\mu_r$&[Pa]  & Shear modulus&Y  \\ 
	\hline 
\end{tabular} 
\end{center}
\caption{Biot-Stoll parameters and their effect on the shear wave. The pore radius parameter is not included as the flow in the pores is assumed to be laminar. The right-hand column shows whether the parameter affects the shear wave or not}
\label{table:BiotParameters}
\end{table}
\subsection{Redundant parameters in the Biot shear wave model}
The low frequency Biot model  is characterized by the ten parameters shown in Table \ref{table:BiotParameters}. Seven of them affect shear wave propagation as indicated in the right-hand column, but several of them are connected as \eqref{eq:BiotParameters} indicates.

The parameter $\omega_\epsilon$ depends on the mass coupling density $\rho_c$. In addition it depends on the ratio of the viscosity, $\eta$, and the permeability, $B$. They do not influence any of the other parameters $\omega_\sigma$ and $c_0$ so it is clear that it is their ratio which matters. 
Likewise, the sound velocity, $c_0$, depends on the ratio of the shear modulus, $\mu_r$ and the aggregate density, $\rho$. 

Furthermore for the third parameter, $\omega_\sigma$, the denominator in the expression is usually larger than $0.8$ so the ratio of $\omega_\sigma$ and $\omega_\epsilon$ is not very sensitive to changes in the densities, $\rho$, $\rho_c$, or $\rho_f$. That means that neither is it very sensitive to changes in tortuosity, and it will mainly be $\omega_\epsilon$ which depends on $\alpha$ in the ratio $\eta/(\alpha B)$. 

In this way three of the seven parameters can be said to be redundant and the seven parameters from Table \ref{table:BiotParameters} ($\eta$, $B$, $\alpha$, $\rho_f$, $\rho_s$, $\phi$, and $\mu_r$) may be reduced to  four  by combining the three first ones into one, $\eta/(\alpha B)$. Alternatively, the four parameters may be stated as $\eta/\alpha$, $\rho_c$, $\rho$, and $\mu_r$.
In case one wants to compute $\omega_\sigma$ exactly, a fifth parameter, $\rho_f$, also needs to be included.

In this way the number of parameters in the Biot shear model with some approximation is four as in the Zener medium model, and five in the exact case.

%

\subsection{Extension to poroviscoelasticity}

In the sediment acoustics field, the Biot model has been amended in order to extend the model from a rigid porous frame, like a porous rock or bone, to one where the grains are allowed to move. This is a model for a saturated sediment and it is not unlikely that it may be more appropriate for tissue than the rigid frame implied in the Biot model also. 

In this case viscosity is introduced in the solid frame in addition to in the flowing liquid. The model is called  the Biot squirt flow and viscous drag (BICSQS) model  \cite{chotiros2004broadband}. This viscosity is added by allowing a relaxation model for the shear modulus, thus allowing for a frequency dependent complex shear modulus or a dynamic shear modulus,  described by a characteristic frequency $\omega_\mu$:
\begin{equation}
\mu' = \mu \left(1+ i \frac{\omega}{\omega_\mu} \right)
\label{eq:viscousChotiros}
\end{equation}
When the relaxation is included in \eqref{eq:BiotShear-dynamicModulus} and \eqref{eq:Zener-dynamicModulus}, the result is
\begin{equation}
E(\omega)  
=  \mu \frac{(1+ i \omega/\omega_\mu)(1+i \omega/\omega_\epsilon)}{1+i\omega/\omega_\sigma}
\label{eq:BICSQS-shear-modulus}
\end{equation}

\begin{figure}[bt]
\centering
\includegraphics[width=0.8\columnwidth]{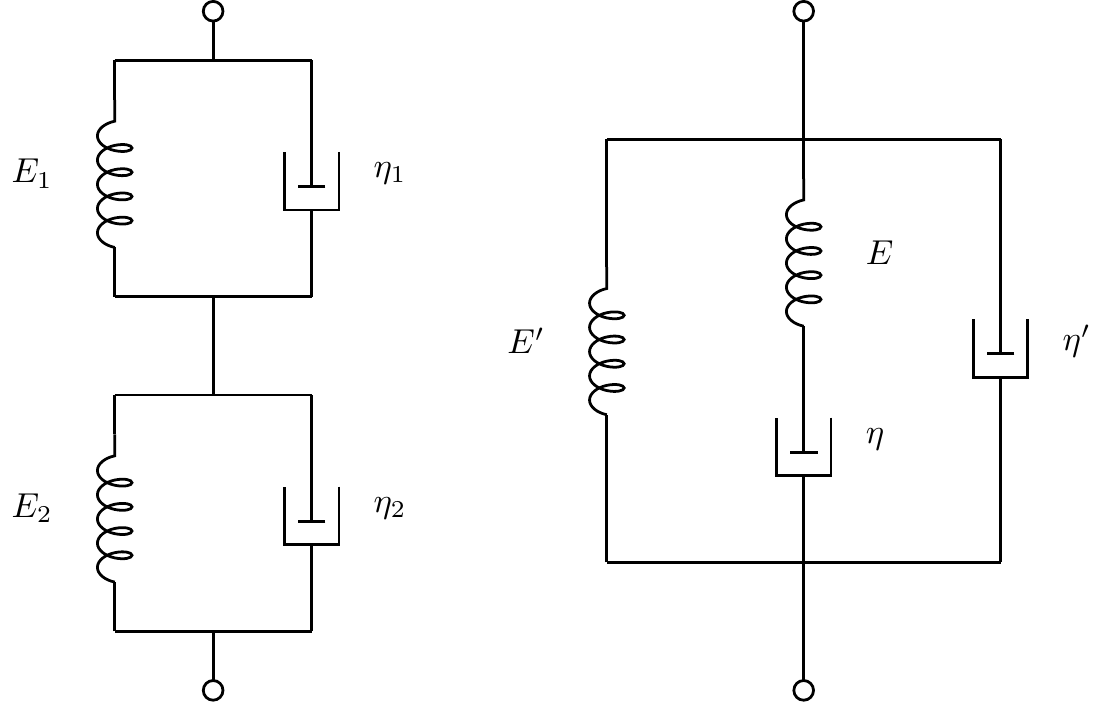}
\caption{The equivalent to the shear wave poroviscoelastic model called  the non-standard four-parameter model in \cite{tschoegl1989phenomenological} (left) with its conjugate to the right}.
\label{fig:NonstandardFour1}
\end{figure}

In the limit as frequency goes to zero the dynamic modulus approaches $E(0)= \mu$ and as the frequency approaches infinity it approaches $E(\infty) = \infty$.
This model is called the 
non-standard Four-parameter model in \cite{tschoegl1989phenomenological} (Chap.~3.4.2). Its spring--damper realization is shown in Fig.~\ref{fig:NonstandardFour1} with the equivalent conjugate model to the right. Therefore, even in this case, an equivalent viscoelastic model may be found.

\section{Discussion and Conclusion}
The concept of the time- and frequency-spectral decompositions of viscoelastic systems has been developed and applied to the fractal Zener and Kelvin-Voigt models. This summation of multiple elementary models 
is an idea which is independent of the fractional models. In fact it is the idea behind the relaxation spectrum models of Kelvin and Wiechert dating from 1888 and 1893 respectively \cite{markovitz1977boltzmann}.
In addition to the coverage in \cite{Mainardi2010}, the book \cite{tschoegl1989phenomenological} devotes several sections to it (Chap.~3.5-3.6). Similar ideas of model fitting have been used in biomechanics and elastography also. In the white matter model of \cite{cheng2007unconfined} the shear relaxation modulus was for instance modeled with three very slow relaxation terms (0.01 Hz and slower) and a similar model with two terms was  used in \cite{caenen2015versatile}.

The idea behind the particular weightings implied in the time-spectral and frequency-spectral functions found here is however to make the sum approximate the behavior of the fractional models, i.e.~in terms of power laws in the frequency domain. Therefore it parallels the modeling of  arbitrary power law attenuation in medical ultrasound over a limited frequency range with a few terms in \cite{Yang2005}. 

The surprising result is that the weighting has a long-tailed distribution. This is a property which is  associated with a fractal geometry. Here it characterizes the distribution of individual relaxation processes both in the time and the frequency domains. An interesting  special case is the hysteresis model recently proposed by \cite{parker2015could} and this could possibly shed some light on the interpretation of the distribution. The fractal distribution is an explanation that seemingly is based on a completely different physical mechanism than the link shown between the fractional damper and a linearly time-varying viscosity in \cite{pandey2016linking}. But perhaps the link is even deeper and suggests that such a time-varying viscosity in some way is associated with a medium with a fractal distribution of relaxation processes?

The connection between the fractal distribution and the fractional constitutive equation of an absorbing medium is also a result that complements recent results for the alternative attenuation mechanism due to scattering rather than absorption. An example is \cite{fouque2007wave, lambert2015bridging} where it is shown that a fractal distribution of random scatterers  leads to attenuation that varies with frequency according to a power law. The fractal -- fractional derivative connection can also be compared to the geometrical and physical interpretation of  fractional integration and differentiation in \cite{podlubny2002geometric}, but the result here is more specific to one particular application, namely viscoelasticity. 

The poroelastic model depends on ten independent parameters, of which seven influence the shear wave solution, and seems to build on an entirely different foundation than the viscoelastic models. Despite that it has been shown to be equivalent to a Zener model. This was done by comparing dispersion relations, and not unsurprisingly the result is the same whether the dispersion relations from the original Biot theory or those from the Biot-Stoll theory are analyzed. 
This result is an extension of \cite{bardet1992viscoelastic} 
where it was 
shown that in the low loss/low frequency approximation, an equivalent can be found between the poroelastic model and a Kelvin-Voigt model. That was done by comparing approximate expressions for the velocity and attenuation. 
As the Zener model in the low loss/low frequency case is equivalent to the Kelvin-Voigt model \cite{Holm2011}, the result found here also agrees with that result. 

It is also shown that the seven parameters of the poroelastic model can be reduced to five, and even four as in the Zener model if a small approximation is allowed. When the Biot model is extended to include viscosity in the frame as in \cite{chotiros2004broadband}, an extra damper has to be added to the Zener model making it into what is called the non-standard four-parameter model in \cite{tschoegl1989phenomenological}.


\end{document}